# Concept-Oriented Programming

Alexandr Savinov

http://conceptoriented.org/savinov

**Abstract.** Object-oriented programming (OOP) is aimed at describing the structure and behaviour of objects by hiding the mechanism of their representation and access in primitive references. In this article we describe an approach, called concept-oriented programming (COP), which focuses on modelling references assuming that they also possess application-specific structure and behaviour accounting for a great deal or even most of the overall program complexity. References in COP are completely legalized and get the same status as objects while the functions are distributed among both objects and references. In order to support this design we introduce a new programming construct, called concept, which generalizes conventional classes and concept inclusion relation generalizing class inheritance. The main advantage of COP is that it allows programmers to describe two sides of any program: explicitly used functions of objects and intermediate functionality of references having cross-cutting nature and executed implicitly behind the scenes during object access.

# 1 Introduction and Motivation

References always have been used in object-oriented programming (OOP) as a mechanism of object identification and access. However, OOP does not provide any means for modelling references. They are supposed to be primitive elements of the programming model, i.e., references need to be provided by the program translator and are beyond the scope of OOP. When writing an object-oriented program we focus on describing behaviour of entities by using classes, inheritance, method overriding and other mechanisms rather than on how these entities will be represented and accessed. In this sense OOP is primarily aimed at entity modelling as opposed to identity modelling, i.e., OOP provides excellent means for modelling the structure and behaviour of objects by delegating the problem of their identification to the translator which determines how each concrete object should be represented and accessed.

In fact, such an explicit concentration on describing entities (objects) is not an ignorance of identities but rather can be characterized as complete abstraction from identity modelling and from the problems arising at the level of object representation and access. Such abstraction provides significant advantages over the approaches where we have to bother not only how an object behaves but also how it has to be identified. The thing is that entity modelling and identity modelling are different concerns which have to be somehow separated so that one and the same code does not mix these two types of functionality. The solution provided by OOP for such a separation consists in abstracting from identity modelling and focusing on entity modelling (the behaviour of objects). Such a code is much cleaner, easier to write and maintain, and more reliable because behaviour of objects does not depend on the way they are represented and accessed. One compiler can generate references in one format while another compiler can use other references and access mechanisms which are more suitable for the platform and run-time environment. And in all these cases objects are guaranteed to retain the desired functionality independent of their references. For example, if we defined a class for describing bank accounts with a business method for crediting this account then its behaviour is independent of the format of references, memory allocation mechanism and the procedures executed during access which all are generated by the compiler.

Yet, using this abstraction from the level of identity modelling and focusing only on entity modelling has also significant disadvantages. The main problem is that the translator responsible for providing references and other functions specific for identity modelling is completely unaware of the problem domain and goals of this concrete program and its objects. The translator knows nothing about the purpose of the program objects and their specific requirements to the identification and access mechanism. In this case the programmer can rely on one universal representation and access mechanism which is normally based on some kind of memory manager provided by the operating system. In this case abstracting from references has a negative consequence of loosing any possibility to control and influence how objects are represented and access. In the case of relatively small standard tasks it is not a problem because platform-specific references provided by the translator are quite efficient and allow the programmer to concentrate on business logic of the objects. However, for larger programs we can observe an interesting phenomenon: more and more code deals with object identification by describing how they are represented and what happens when they are being accessed. In large programs this type of code is already everywhere and spans the whole program so that functions executed during access account for the most of the system complexity. Artificial application-specific object identifiers, primary keys from persistent storage, remote references, container-specific handles and numerous domain-specific identities like bank account numbers are only a few examples of such non-primitive references. And each user-defined reference in the program needs the corresponding access mechanism like loading/storing object state given its primary key or serializing/deserializing when accessing an object on its remote reference. Thus the hidden world of identity modelling breaks into the world of object-oriented programming where its tasks have to be solved using only old methods of entity modelling provided by OOP.

Of course, there exist numerous patterns and techniques that can be used to model references using only methods of OOP. For example, we can use proxies, smart pointers, delegation, identifier keys or interception to name only a few such mechanisms. However, all these and many other mechanisms still assume that objects and entity modelling provide a fundamental basis of programming while references are a kind of second-class citizens in this entity world. In other words, references with representation and access functionality are supposed to be important but still secondary elements of the programming model. As a consequence, references can be and should be modelled but in OOP we can use only classes which are not suitable for this task. For example, we might define a new class field which stores this object identifier like bank account number. Or, we might introduce a class which describes the structure of remote reference and so on. But the problem is that they are still normal classes and the translator is completely unaware that we have actually defined a new identifier. The translator does not know that the programmer tries to do its work by extending primitive references and their built-in identity modelling functionality. Indeed, the real intention of the programmer consists in introducing domain-specific identifiers instead of or in addition to the platform-specific references. Yet the language and translator cannot help him in this task and still treat all constructs in terms of OOP. The situation is analogous to using object-oriented approach in a procedure-oriented programming language where the programmer has to manually implement all the necessary OOP mechanisms and follow its conventions, for example, assuming that the first parameter of any procedure is the object reference while virtual functions are manually implemented using switch statement.

The main problem of implementing references and their functions using conventional object-oriented constructs is that the translator does not constrain the programmer and cannot provide its support. Thus the programmer will "creatively" implement various representation and access functions in such a way that it will be very difficult for other programmers to understand what role is played by different classes. It is the same as determining whether the first parameter of a procedure is the target object identifier or simply a normal parameter or whether some switch statement implements a virtual function or has some other purpose. A more serious problem is that manual implementation can be really very difficult without the support from the language and translator. In particular, the fundamental problem is that the representation and access functions associated with references have a cross-cutting nature, i.e., they are spread all over the program and in this sense they are analogous to aspects in the aspect-oriented programming. For example, one type of reference can be used to represent many classes of objects and then these functions have to be called when any of the target objects is about to be accessed. Since OOP is known to have no good support for cross-cutting functions [15], it is difficult to implement references in it. The third problem is that functions of references have to be activated implicitly behind the scenes during object access rather than called explicitly. For example, whenever a bank account is credited it is necessary to execute some procedure which is specific for its type of reference. However, the problem is that we do not want to call these functions explicitly because we lose separation of concerns and get mixed code. Instead, we want our translator to understand that for this type of method calls some other functions have to be triggered automatically.

In order to overcome these and other problems we propose a new approach to programming, called concept-oriented programming (COP), which focuses on identity modelling by providing mechanisms and constructs for describing representation and access functions. However, instead of proposing some additional mechanisms that can help in implementing references we start from changing the fundamental principles. In particular, one of the main assumptions of COP is that references and identity modelling play a primary role in any system which is different from and opposed to the role of objects and entity modelling. In particular, we can well think of a program without entities and business logic but having a rather complex functionality encapsulated in references. Moreover, a program can consist of only references and have arbitrary complexity but it cannot exist without references because they provide the only way for object access. This assumption makes references first-class citizens in the world of computer programming, i.e., references are completely legalized and get the same status as objects. In particular, just as objects, references possess structure and behaviour. In this sense, if the main goal of OOP consists in describing behaviour of objects then the main goal of COP consists in describing both objects and references.

COP divides the whole programming model into two areas: identity modelling and entity modelling. However, they cannot exist in isolation and need to be modelled in inseparable unity. In particular, references and objects (elements of the identity and entity worlds, respectively) are two sides or flavours of one thing. In other words, any element is supposed to have two sides: one identity and one entity. OOP provides effective means for entity modelling while COP adds new mechanisms for identity modelling and combines them together into one general approach. The duality of references and objects in COP creates a nice yin-yang style of balance and symmetry between two sides of one reality having a cross-cutting nature. Thus we still manipulate elements as one whole but have a possibility to separate different functions using the duality principle. Very informally, this could be compared with the introduction of complex numbers in mathematics which have two constituents: a real part and an imaginary part. Yet these two constituents are always manipulated as one whole and this makes mathematical expressions much simpler and more natural. The same effect we get in programming when separate references and objects but manipulate them as one whole: program code gets simpler and more natural because elements can exhibit properties of both references and objects. In this context, it is important to emphasize that the main goal of COP is not to mechanically add mechanisms for identity modelling but rather to integrate them with OOP by generalizing notions where necessary and producing synergy effect and paradigm shift.

One important consequence of legalizing references and bringing identity modelling into the area of computer programming is the possibility to model indirection and virtualization. Indeed, reference is simply a convention for identifying objects which has no direct relation to the real object location. For example, city or street name do not have any information on how to find the represented object. However, in COP, references possess not only structure but also behaviour and hence they are active elements of the program. A reference in this sense is a virtual identifier which knows how to find the target entity. The functionality associated with a reference is automatically triggered during object access and in large programs these hidden implicitly activated functions can account for most of the system complexity. Thus object representation and access functions can be thought of as "dark matter" of any system which is invisible but plays a very important role in its functioning. Instead, this layer of functionality is activated implicitly during object interaction. In this sense the goal of COP consists in making this hidden layer of functionality explicit by providing means for modelling this dark matter and its behaviour. Informally, OOP can be thought of as using the principle of instantaneous action dominating in physics of 18$^{th}$ century. In contrast, COP assumes that any interaction needs some environment to propagate and this environment is responsible for many important functions of the program. Elements of the concept-oriented program are represented by virtual identifiers and access to them is always indirect, i.e., some intermediate functionality is activated for propagation of method calls. It can be compared to using virtual memory in contemporary processors where each access is performed via translation of the virtual address into a physical address. Essentially, the programmer in COP can implement arbitrary virtual address systems for the program objects with the necessary access procedures and reference resolution mechanisms. Note again, that we assume that identity modelling is part of any problem domain and hence each program reflects the structure of identities and access mechanisms which are specific to this concrete application.

The main goals of COP can be summarized as follows:

- Provide an explicit and strong notion of *identity* (reference) in programming languages where references are active elements with their own structure and behaviour
- Providing means of *access indirection* so that the programmer can describe what happens during object access
- Provide a mechanism for modelling domain-specific *address spaces* and containers which can be then used for managing objects
- Provide a mechanism for modularizing *cross-cutting concerns*
- Making programming *set-oriented* where any element is not simply an instance among other instances but rather is a set of other instances

Importantly, all these goals should be met using a minimum number of core notions and mechanisms, ideally, by relying on only one primary construct – class (possibly generalizing and modifying its roles).

The paper has the following layout. Section 2 is devoted to describing concepts and concept inclusion relation. In particular, Section 2.1 describes what we mean by concepts and what their main purpose is. Section 2.2 introduces inclusion relation and describes how it can be used to model hierarchical address space and complex references. The sequence of access in the hierarchical address space is described in Section 2.3 while Sections 2.4 and 2.5 show how inheritance and polymorphism change in COP. References have to be somehow bound to the represented objects and this mechanism is described in Section 3. In particular, Section 3.1 defines a special method of concepts which is responsible for finding the represented object. Section 3.2 describes the sequence of complex reference resolution and how this procedure can be optimized using a special data structure called context stack. In Section 3.3 we describe how concept instances are created and deleted. Several useful operations for manipulating references are defined in Section 4. Section 5 is a short overview of related work and Section 6 makes concluding remarks. In diagrams throughout this paper, we will use a convention that objects are shown as (white) rectangles while references are drawn as (grey) rounded rectangles. Code snippets demonstrating various COP features and mechanisms will use Java-like syntax. In these examples we assume that return is a special variable which stores a value which is returned at the end of the method scope.

## 2 Concepts and Concept Inclusion

### 2.1 Concept as a Language Construct

COP postulates that an element of a program has two sides: identity and entity (the principle of duality). To describe their structure and behaviour we can use conventional classes. What is new in our approach is that COP uses two sorts of classes for describing two parts of an element. The class that is used to describe structure and functions of identities is called a *reference class*. The class that describes structure and functions of entities is referred to as an *object class*.

Object classes are equivalent to normal classes as they are defined and used in OOP and their instances are referred to as *objects*. Thus if no reference classes are defined then we get an object-oriented program which consists of objects represented by primitive references. Reference classes are used to describe object representatives and their instances are referred to as *references*. In the presence of reference classes a program consists of two types of elements – objects and references – and the program functionality is distributed among them. Thus references are as important as objects and the task of system design consists in determining how different functions have to be distributed between reference classes and object classes.

For example, let us assume that we need to model bank accounts. An account consists of its identity and its entity which are modelled by two classes as shown in Listing 1.

Listing 1. Reference class and object class defined separately.

```
01  reference AccountReference { // Reference class
02    char[10] accNo; // Identifying field
03    ... // Other members of the reference class
04  }
05
06  object AccountObject { // Object class
07    double balance; // Entity state field
08    ... // Other members of the object class
09  }
```

Here we use the keyword 'reference' to mark a class as a reference class and the keyword 'object' to mark a class as an object class. We might also add other members to these classes, say, opening date field or a method for getting balance but the main idea is that the language has an explicit indication for the purpose of any class which is necessary to organize the appropriate sequence of access.

These two classes have different names and they can be separately instantiated in the source code. The main goal of introducing a reference class consists in using *custom references* for representing objects. Therefore it is necessary to provide two class names for each new variable where one class specifies the object type and the other class specifies its reference type. For example, using the style of the Transframe programming language [31] we could declare a variable referencing an account object using our custom reference as follows:

```
AccountObject account of AccountReference;
```

This declaration means that variable `account` will contain a reference of class `AccountReference` which represents an object of class `AccountObject`. The same can be written using the style of C++ smart pointers [34] as follows:

```
AccountReference<AccountObject> account;
```

Again, we declare a variable which contains a reference of one class pointing to an object of another class. In any case it is important that the two classes are defined independently and we need to specify both of them when declaring new variables, parameters, fields or return values. In particular, one reference type can be used for representing different objects and one object can be represented by many different reference types.

Using *separately* reference classes and object classes is possible but is not really fruitful and therefore is *not* used in COP. In this case references and objects are defined separately and there is no connection between them because these classes are coupled only when they are used for declaring a variable. Thus the classes describing one element cannot use each other. Essentially, this use of reference and object classes does not satisfy the principle of duality which postulates the inseparable *unity* of two parts of one thing. To overcome this problem we propose to use a new programming construct, called *concept*, which combines the two classes.

**Definition 1 [Concept].** *Concept* is a pair consisting of one reference class and one object class

Within a concept, reference class and object class lose their independence and can exist only as its parts. If we need to have only a reference class then it can be defined as a concept with the empty object class. And if we need to have only an object class then it can be defined as a concept with the empty reference class. The latter is equivalent to normal classes. Thus it is important to understand that in COP the programmer manipulates only concepts, i.e., pairs of classes rather than individual classes.

Listing 2. An example of concept.

```
01  concept Account // One name for the pair of two classes
02    reference { // Reference class of the concept
03      char[10] accNo;
04      ... // Other members of the reference class
05    }
06    object { // Object class of the concept
07      double balance;
08      ... // Other members of the object class
09    }
```

For example, instead of defining separately one account class and one account reference class (Listing 1) we now define one concept which contains the both (Listing 2). Every concept definition begins with the keyword 'concept' followed by a concept name. Reference class and object class are still marked by the keywords 'reference' and 'object', respectively.

Since reference classes and object classes cannot be defined independently, they cannot be used separately in the program. Instead, we need to use concepts only, i.e., pairs of one reference class and one object class. Thus concepts in COP are used where classes are used in OOP when declaring types of variables, fields, parameters, return values etc. For example, let us consider the following code:

```
Account account = getAccount();
Person person = account.getOwner();
Address address = person.getAddress();
```

Strictly speaking, it is not possible to determine if it is an OOP program or a COP program from this fragment because we do not know how its types (`Account, Person, Address`) are defined. If they are defined as normal classes then it is an OOP program. Otherwise, if they are defined as concepts (for example, as concept `Account` in Listing 2) then it is a COP program. So we can still write a program in object-oriented style by declaring variables and calling methods with the only difference that concepts are used instead of classes.

Then the question is what really changes if we replace classes by concepts? The main consequence of using concepts is that variables contain custom references in the format defined by the concept reference class. In contrast, if a variable were defined using a class then it would contain a primitive reference chosen by the compiler. Thus changing concept definition we can effectively influence what is actually stored in the variables in the program and passed in the parameters. But why do we need to represent program objects by custom references? The answer is that we want to represent entities by their natural identities as they are used in the problem domain rather than by system-specific surrogates. In addition, these application-specific references encapsulate quite complex functionality which cross-cuts the whole program and is activated implicitly during object access. It is also important to understand that custom references defined by concepts have wider scope rather than primitive references with fixed scope, i.e., we can store such references or pass them to a remote context.

For example, if `Account` is a concept defined in Listing 2 then variable `account` will store account number as its content. This account number is supposed to *indirectly* represent some account object and it will be passed to method parameters, stored in object fields or returned from methods as if it were normal (primitive) object reference. Analogically, if `Person` is a concept then variable `person` would store passport number and birth date which both indirectly represent the owner of the account. Note that these references can be stored or passed to programs or computers and they are still valid because of the inherent indirection.

One interesting consequence of introducing concepts is that now one and the same method can be defined twice, i.e., reference class provides a definition of a method and object class has its own definition of this same method.

**Definition 2 [Dual methods].** Methods with the same signature defined in both reference class and object class of a concept are called *dual methods*.

The definition provided in the reference class is called a *reference method* while the definition provided in the object class is called an *object method*. Although one of them can be absent we will assume that both definitions are always available and have a default implementation if not defined explicitly by the programmer.

The main issue with dual methods is that in source code there is no indication which of two definitions to use. In other words, methods are used as usual by specifying its name and parameters but if it is applied to a concept instance then this concept may provide two definitions for this method and then the question is which one to really call. For example, if method `getBalance` is defined in both reference class and object class of concept `Account` and we apply it to variable `account` of this concept (Listing 3, line 18) then what definition has to be actually executed?

In order to resolve this issue COP uses the following principle:

**Principle 1** [Precedence of dual methods]. Reference methods of a concept have precedence over its object methods.

Listing 3. Precedence of reference methods.

```
01  concept Account
02    reference {
03      char[10] accNo;
04      double getBalance() { // Reference method
05        print("=== Account::getBalance reference method");
06        return = 0;
07      }
08    }
09    object {
10      double balance;
11      double getBalance() { // Object method
12        print("--- Account::getBalance object method");
13        return = balance;
14      }
15    }
16
17  Account account = getAccount();
18  double balance = account.getBalance();
19
20  $ === Account::getBalance reference method
```

This means that when a method is applied to a variable then the reference method of its concept will be called. For example, the statement `account.getBalance()` will use the definition provided in the reference class of concept `Account` (Listing 3, line 20). Essentially, we can act only on what is stored in variables by value, i.e., we work with references but assume that they should affect the represented objects by somehow propagating the request. By calling method `getBalance` we want to access the account object however this operation cannot be executed directly because of the absence of primitive reference. Indeed, the account can reside anywhere in the world and direct access is impossible. In this case the reference stored in the variable intercepts this call and executes all the necessary intermediate actions. Thus the reference is the only element that knows where the object is located and how to access it, particularly, how to call its methods. Another role of references is that they can protect objects from direct access and encapsulate functionality that has to be executed before any object method. This mechanism creates also the illusion of instantaneous action by allowing the programmer to control the actions executed implicitly during object access. Indeed, we simply call a method without the need to know the peculiarities of object access procedure. What is even more important, such a code is easier to maintain because the use of object methods (business logic) is separated from the logic of representation and access.

When concept methods are used we do not distinguish between reference methods and object methods. However, they need to be distinguished when used from within the concept. For example, if a reference method got control then normally it has to call its object methods. In order to distinguish them, a programming language needs some syntactic means. We will use a simple convention that this reference is denoted by the keyword 'reference' while this object is denoted by the keyword 'object', which is analogous to 'this' or 'self' keywords in OOP. Using these two keywords we can distinguish between reference and object members. For example, `reference.getBalance()` is a call of the reference method while `object.getBalance()` is a call of the object method. The process of moving from reference to the represented object is referred to as *meta-transition*. (It is a modified version of the term borrowed from the Metasystem Transition Theory developed in cybernetics.) This process will be studied in detail in Section 3.

Listing 4 provides an example where a reference method calls the dual object method and the corresponding sequence of access is shown in Fig. 1. If this method is applied to an account variable then first the reference intercepts this call and executes the necessary intermediate operations (step 1 in Fig. 1). Then it calls the dual method using the 'object' keyword to indicate that it is an object method (step 2). At this point the process performs meta-transition and moves to the object world (step 3). And finally the object method executes the necessary operations with this object and returns some result (step 4). The output of this method call is shown at the end of Listing 4.

Listing 4. Call of object method.

```
01    concept Account
02      reference {
03        char[10] accNo;
04        double getBalance() {
05          print("=> Account::getBalance reference method");
06          balance = object.getBalance(); // Object method is invoked
07          print("<= Account::getBalance reference method");
08          return = balance;
09        }
10      }
11      object {
12        double balance;
13        double getBalance() {
14          print("-> Account::getBalance object method");
15          return = balance;
16          print("<- Account::getBalance object method");
17        }
18      }
19
20    Account account = getAccount();
21    double balance = account.getBalance();
22
23    $ => Account::getBalance reference method
24    $ -> Account::getBalance object method
25    $ <- Account::getBalance object method
26    $ <= Account::getBalance reference method
```

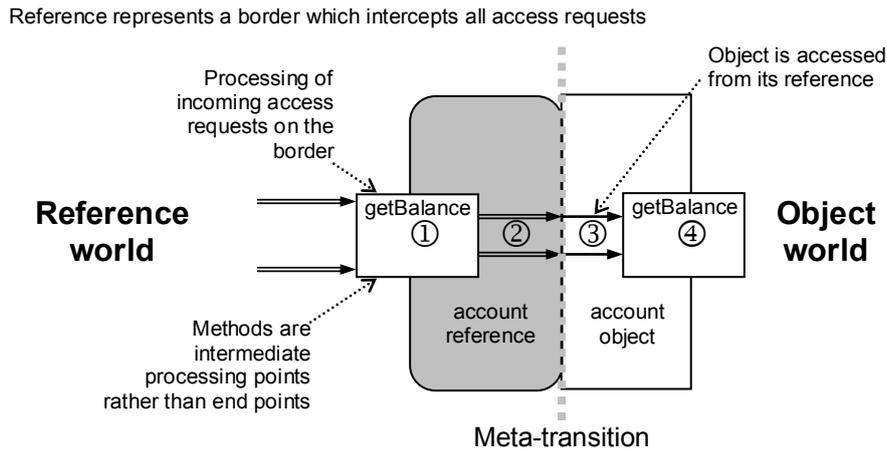

Figure 1. Reference intercepts accesses to the object.

## 2.2 Concept Inclusion and Complex References

Just as classes use inheritance relation, concepts use *inclusion relation* to establish a hierarchy where each concept has a parent concept and is said to be included in it. In source code, concept inclusion will be declared using the keyword 'in' followed by a parent concept name. For example, if concept `SavingsAccount` is included in concept `Account` then it is written as follows:

```
concept SavingsAccount in Account
  reference { ... } // Reference class
  object { ... } // Object class
```

If parent concept is not specified explicitly then by default it is assumed to be the root concept in the inclusion hierarchy.

The main difference between inclusion and inheritance is that inclusion is interpreted as IS-IN relation while inheritance is interpreted as IS-A relation. This means that just as concepts themselves, concept instances exist in a hierarchy at run-time where one parent instance may have many child instances. For example, one account (an instance of concept `Account`) may have many savings accounts (instances of concept `SavingsAccount` which is included in `Account`). In OOP terms, this means that an object may have many extensions or, vice versa, extensions can share one base object. References and objects consist of several parts, called *segments*, where each part is an instance of one concept in the inclusion hierarchy.

Since objects exist in a hierarchy their references provide only a relative identifier with respect to the parent object, i.e., objects belonging to different parents may well have the same reference. For example, main accounts may have savings accounts with the same number or different cities may have the same streets. In order to identify an object in the inclusion hierarchy it is necessary to use not only its own reference segment but also the parent reference segments.

**Definition 3 [Complex reference].** *Complex reference* is a sequence of reference segments where concept of each next segment is included in the concept of the previous segment. A sequence of objects represented by a complex reference is referred to as a *complex object*.

A complex reference which starts from the root is called a *fully qualified reference* or global reference. Otherwise, if a reference starts from some internal concept then it is called a *local reference*. Complex references are represented as one value, i.e., its segments are passed and stored side-by-side (like object segments in OOP). However, object segments have their own references and hence can be stored in different places.

If a concept has a parent concept and then it is used as a type of some variable then this variable will contain a complex reference which starts from some parent concept of this type (root by default) and ends with this concept or one of its children (how reference length can be controlled is described in Section 4). For example, a variable of concept `SavingsAccount` will contain two reference segments: the first is defined in concept `Account` and the second defined in concept `SavingsAccount` (Fig. 2). A savings account is then indirectly represented by two numbers: the main account number and the sub-account number identifying its object within the main account.

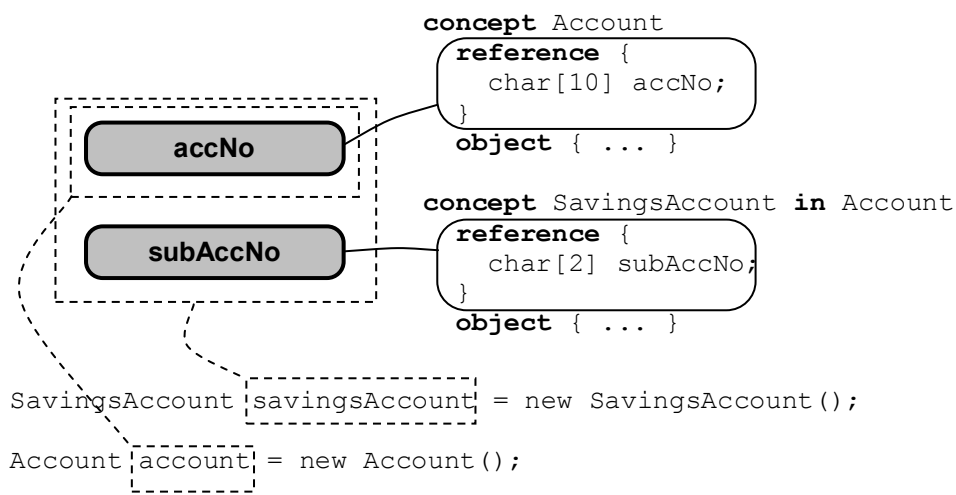

Figure 2. Structure of complex reference.

As already noted, segments of a complex reference are stored and passed together as one data structure. It is analogous to how objects in OOP are normally stored in one chunk of memory one segment next to the other. In this sense inclusion relation for reference classes can be thought of as a normal inheritance. The difference from objects in OOP is that references need not to start from the root because frequently it is enough to have a local reference which can start from any concept. For example, local computations within one account can use a relative savings account number assuming that it is an internal sub-account. However, by default references will be generated as global references. For example, if `SavingsAccount` is included in `Account` which is included in `Bank` concept then a reference to a savings account object will consist of three segments as shown in Fig. 3. These three reference segment represent three object segments which constitute one complex object. Note that the object segments need not to reside next to each other. For navigation over inclusion hierarchy, two keywords are used: 'super' refers to the parent element and 'sub' refers to the child element.

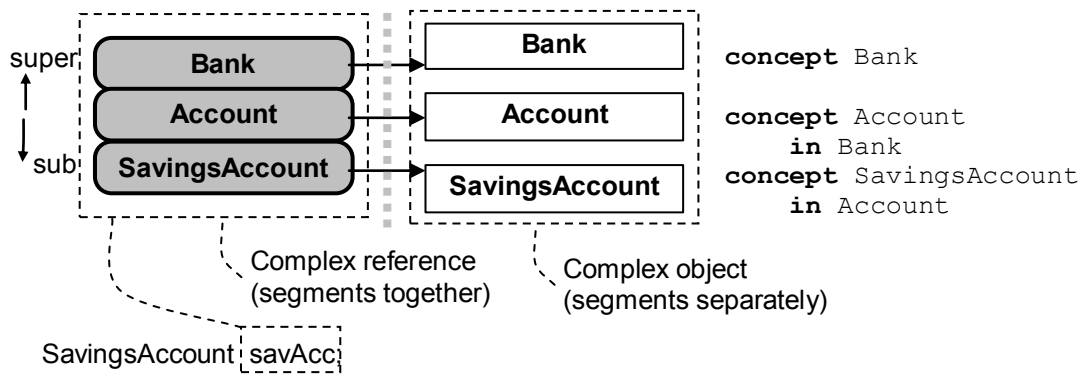

Figure 3. Complex reference and complex object.

### 2.3 Sequence of Access

Earlier in Principle 1 (Section 2.1) we postulated that references intercept all accesses to the represented object and object members can be accessed only from within a reference method. However, if variable stores a complex reference then every one of its segments can implement the same method. For example, both concepts `Account` and `SavingsAccount` can implement method `getBalance` in their reference class. Then the question is which of these two methods has to be executed if it is applied to an instance of concept `SavingsAccount`? In other words, the problem is to determine which reference segment should process all incoming access requests. In order to resolve this ambiguity we use the following principle:

**Principle 2** [Reference method overriding]. Parent reference methods have precedence over (override) child reference methods.

In other words, parent reference methods of higher segments intercept any access to the child reference methods of lower segments. In our example, `getBalance` of `Account` reference will be called first and only after that it is possible to call `getBalance` of `SavingsAccount` reference.

This principle is quite natural and simply reflects the fact that any attempt to enter a space must be intercepted at the border (Fig. 4, left). Higher segments represent external spaces while lower segments represent internal spaces. In order to reach an element we always start from the external space and then proceed by entering narrower scopes. At each intermediate border the request is intercepted by the method having the same signature and the programmer can perform the necessary actions. (Of course, if this processing is not needed then the interception can be optimized, i.e., if the reference class of the parent concept does not define the method then it will not be intercepted and then the child method can be called directly.) Such a sequence of access effectively means that the only possibility to access an object consists in intersecting all the intermediate borders that separate it from the outside world.

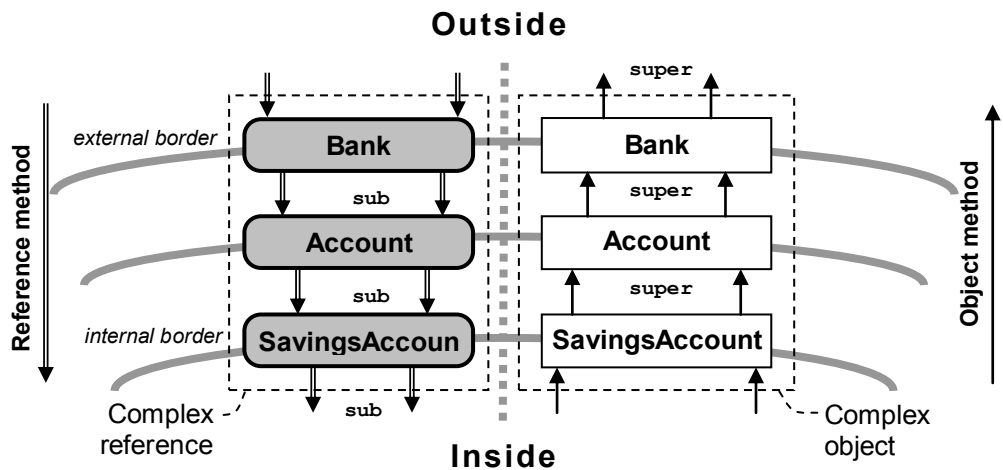

Figure 4. Duality of method overriding.

An important application of this principle is the mechanism of method overriding. However, the direction of such overriding is opposite to the conventional one (as used in OOP), which means that parent reference methods override child reference methods. (Note that this principle belongs to reference methods only.) It is a dual form of the mechanism of method overriding in OOP, which allows a sub-class to provide a more specific implementation of a method defined in its super-class.

Listing 5. Reference method overriding.

```
01  concept Bank
02    reference {
03      void doSomething() {
04        print("=> Bank: enter bank scope");
05        sub.doSomething(); // Go inside
06        print("<= Bank: exit bank scope");
07      }
08      ...
09    }
10    object { ... }
11
12  concept Account in Bank
13    reference {
14      void doSomething() {
15        print("  => Account: enter account scope");
16        sub.doSomething();// Go inside
17        print("  <= Account: exit account scope");
18      }
19      ...
20    }
21    object { ... }
22
23  concept SavingsAccount in Account
24    reference {
25      void doSomething() {
26        print("    => SavingsAccount: enter sub-account scope");
27        sub.doSomething();// Go inside
28        print("    <= SavingsAccount: exit sub-account scope");
29      }
30      ...
31    }
32    object { ... }
33
34  SavingsAccount acc = new SavingsAccount();
35  acc.doSomething();
36
37  $ => Bank: enter bank scope
38  $   => Account: enter account scope
39  $     => SavingsAccount: enter sub-account scope
40  $     <= SavingsAccount: exit sub-account scope
41  $   <= Account: exit account scope
42  $ <= Bank: exit bank scope
```

Using this principle it is always possible to override any reference method by defining the same method in the reference class of its parent concept. Thus parent reference methods protect child reference method from direct use from outside. It is quite natural principle because it allows any border to control incoming processes. For example, a live cell has a border which checks and controls anything that tries to come in. The existence of the own border is actually a generic property of any system including physical, live and social ones. And it is quite natural that if a system is included in another system than the only way to access it from outside consists in intersecting the parent system border. In this sense reference methods can be thought of as incoming methods for the space described by this concept. The mechanism of reference methods and the inverse principle of overriding allow us to support this approach in programming languages. Note that this principle of overriding is analogous to that used in the Beta programming language [16, 17, 18] and the mechanism of inner methods [7].

Listing 5 (and Fig. 4, left) provides an example illustrating the mechanism of reference method overriding. Variable `acc` (line 34) contains a complex reference of concept `SavingsAccount` which is included in concept `Account` which is in turn included in concept `Bank`. Reference classes of all the tree concepts define method `doSomething` which simply calls the same method of the child reference using the 'sub' keyword and prints two diagnostic messages around this statement. (It is assumed that if there is not child then access on the 'sub' keyword is equivalent to no-op.) If now we apply this method to the variable `acc` of concept `SavingsAccount` then it will print the output shown in lines 37-42. Here we see that any method call is wrapped into a sequence of reference methods starting from the first parent and ending with the last child.

Now let us assume that after calling some object method the process has found itself in the object world. If we call some method from this context then it could be defined also in parent objects. In this case there is some ambiguity concerning what definition of the object method to use for execution. This is actually the same question that has been asked in the beginning of this section for reference methods. However, the answer has the dual form. Namely, for object methods we adapt the conventional OOP principle with the normal direction of access (Fig. 4, right):

**Principle 3** [Object method overriding]. Child object methods have precedence over (override) parent object methods.

If we call some object method which is implemented in all object classes in the concept inclusion hierarchy then the compiler will use the definition provided by this object class (we say, that this method overrides its parent methods). After that this method can continue by calling its parent methods using the 'super' keyword. Thus child object methods protect parent object methods from direct use from inside. Informally this means that if we need some service or support from an object then the most specific one will be provided first while more general services cannot be accessed directly by internal objects.

An example shown in Listing 6 demonstrates the sequence of access on object methods and the logic of object method overriding. In fact, this program can be produced almost mechanically from the program in Listing 5 by moving method definitions from the reference classes to the object classes and using 'super' instead of 'sub'. We also assume that if no reference method has been defined then its default implementation is to pass control to the child (to more specific element) as shown in the previous example. And if it is the last segment with no child then the object method is called. If method `doSomething` will be applied to a variable of concept `SavingsAccount` then it will produce the output shown in lines 37-42.

Listing 6. Object method overriding.

```
01  concept Bank
02    reference { ... }
03    object {
04      void doSomething() {
05        print("->Bank: enter service");
06        super.doSomething(); // Go deeper
07        print("<-Bank: exit service");
08      }
09      ...
10    }
11
12  concept Account in Bank
13    reference { ... }
14    object {
15      void doSomething() {
16        print("  -> Account: enter service");
17        super.doSomething();// Go deeper
18        print("  <- Account: exit service");
19      }
20      ...
21    }
22
23  concept SavingsAccount in Account
24    reference { ... }
25    object {
26      void doSomething() {
27        print("    -> SavingsAccount: enter service");
28        super.doSomething();// Go deeper
29        print("    <- SavingsAccount: exit service");
30      }
31      ...
32    }
33
34  SavingsAccount acc = new SavingsAccount();
35  acc.doSomething();
36
37  $     -> SavingsAccount: enter service
38  $   -> Account: enter service
39  $ -> Bank: enter service
40  $ <- Bank: exit service
41  $   <- Account: exit service
42  $     <- SavingsAccount: exit service
```

If we combine these two examples (Listings 5 and 6) then the following output will be produced:

```
$ => Bank: enter bank scope
$   => Account: enter account scope
$     => SavingsAccount: enter sub-account scope
$     -> SavingsAccount: enter service
$   -> Account: enter service
$ -> Bank: enter service
$ <- Bank: exit service
$   <- Account: exit service
$     <- SavingsAccount: exit service
$     <= SavingsAccount: exit sub-account scope
$   <= Account: exit account scope
$ <= Bank: exit bank scope
```

According to these two principles any process enters a space via some reference method and then goes down along the inclusion hierarchy to more specific reference segments (Fig. 4, left). Then the process switches to the object world where it changes the direction and goes up to more general object segments (Fig. 4, right). So here it is important to understand that changing the current world (between object world and reference world shown as left and right parts in Fig. 4) entails change of the principle of method overriding and other mechanisms to the dual version. Thus the role and properties of reference methods and object methods are significantly different within inclusion hierarchy. Reference methods are used to enter a scope, i.e., they are executed when an access request needs to come into the space. Object methods are viewed as services intended to be used by internal elements which are already in this space. Actually it is analogous to the natural sequence of access used in real systems and organizations. For example, to get a service from some organization it is necessary to enter its scope and then from inside we can use its internal services.

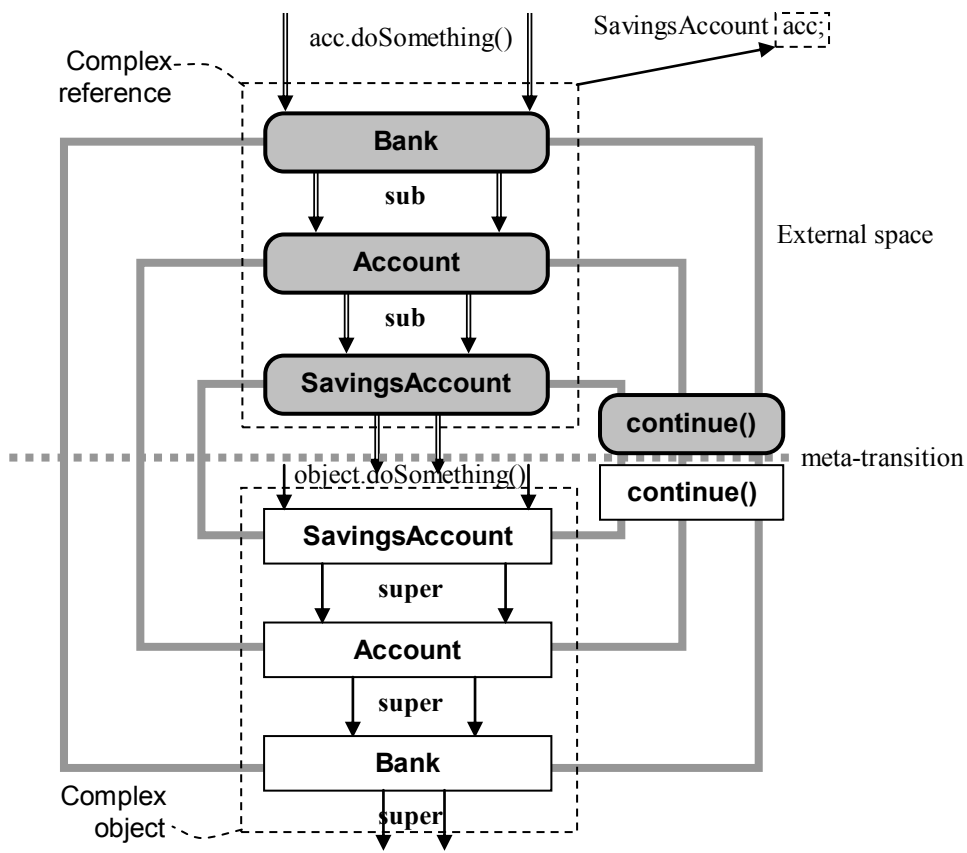

Figure 5. Generic sequence of access.

Diagram shown in Fig. 5 is an alternative view of the generic sequence of access described in this section. Here reference world is shown in the upper part while object world is shown in the lower part. The upper part is what we directly manipulate in the program, for example, by creating a reference of concept `SavingsAccount` consisting of three segments. The three object segments represented by this reference exist in the object world and they can be accessed only using meta-transition (the horizontal line in the middle) implemented in the continuation method (see Section 3). In this example we assume that the process goes down till the last segment where it switches to the object world and then continues its execution via object methods. The access always proceeds in the downward direction however in the middle it changes its semantics.

## 2.4 Inclusion and Inheritance

Earlier we have already noted that inheritance in OOP is interpreted as IS-A relationship. For example, if class `SavingsAccount` inherits class `Account` then any instance of `SavingsAccount` IS-An instance of class `Account`. Or, if class `Button` were defined as inheriting class `Panel` then again each button would be automatically a panel. Here it is important that base and extension cannot be separated, i.e., an extension is always *identified* with its base. In particular, such a complex object is always represented by a single primitive reference and it is not possible to have many extensions for one base. For example, if we create a new savings account then a new base account will be automatically created and if we create a new button then a new panel will be created for it. Inheritance relation is asymmetric because we can reuse (share) base classes at compile-time but we are not able to reuse (share) base instances at run-time. In other words, one base class has many extension classes but one base object has only one extension object. Because of this asymmetry classes exist within a hierarchy while objects exist in flat space where all of them are represented by primitive references of one and the same type.

In contrast, inclusion in COP is interpreted as IS-IN relationship. This means that a new instance of a child concept is created within some instance of the parent concept. In particular, child instances have their own references which distinguish them within their parent instance and it is possible to have many child instances within one parent. Thus base objects and extensions exist independently as normal objects having their own references in the same way as cities, streets and houses are independent objects having their own local identifiers. For example, many savings accounts can be created within one parent account and many buttons can be created within one panel. This makes the whole picture symmetric because both concepts and their instances exist within a hierarchy and the inclusion hierarchy of concepts is used to model the hierarchy of instances. In this case child concepts reuse their parent concept at compile-time while child instances reuse their shared parent instance at run-time. Object segments in COP exist separately at run-time just as their concepts exist separately at compile-time. This significantly changes the role of the parent object. In COP, it is interpreted as a container or environment for many internal elements for which it provides useful services and protects them from direct use from outside.

Although inclusion is different from inheritance, its amazing property is that it generalizes inheritance, i.e., under certain simplifying conditions inclusion is reduced to inheritance where IS-IN is equivalent to IS-A relationship. Such a backward compatibility provides significant benefits by allowing to use conventional approaches developed within such classical disciplines like knowledge engineering (for example, frames [21]) or ontologies [9, 10]. Inclusion can be turned into the classical inheritance if object segments in the hierarchy do not have their identity. This can be done by using empty reference class in child concept (but we still can use reference methods which will intercept incoming access requests). In this case child instances are indistinguishable in the context of their parent object and hence it is assumed that they inherit the parent identity. In other words, parent object and its extensions are considered one and the same thing represented by one (parent) reference. Object allocation in this situation can be optimized by putting them side-by-side in one memory interval. Obviously, here we obtain the case considered in OOP where classes by definition are not able to model identity which is provided by the system. As a consequence all object segments have the same identity in the form of primitive reference.

## 2.5 Inclusion and Polymorphism

Polymorphism is a mechanism which allows a programmer to manipulate objects of more specific types as if they were of the base type. In other words, we can view objects as having the base type without the need to know their concrete more specific type. For example, if we declare a variable as having type `Account` then polymorphism allows us to apply method `getBalance` to references stored in this variable even if they represent more specific account like savings account or checking account and the behaviour has to correspond to the real type rather than to the declared variable type. Let us consider the following example:

```
Account account;
double balance;
account = getSavingsAccount();
balance = account.getBalance(); // Balance of savings account
account = getCheckingAccount();
balance = account.getBalance(); // Balance of checking account
account = getMainAccount();
balance = account.getBalance(); // Balance of main account
```

Here it is enough to know that the object is of class `Account` has method `getBalance`. The essence of polymorphism is that this method is not a concrete procedure known at compile-time but rather a placeholder or label for some general action. In other words, by applying this method at compile-time we actually do not know what will happen at run-time. The real procedure that will be executed at run-time depends on the real type of the object which can be determined only at run-time. In the case of polymorphism, such method invocations are intrinsically indirect. For example, if the real object is a savings account then the balance is calculated using one procedure while for checking accounts it is calculated using another procedure but this procedure can be determined only at run-time.

One approach to implementing polymorphic method calls consists in checking the real object type at run-time and then making a decision what procedure to execute. For this purpose each object has to store information on its real type in some well known field. When a method is applied to an object, the compiler makes an indirect method call using this field for dispatching. The real procedure applied to the object is not known at compile-time and depends on the real object type at run-time. One property of this classical approach to polymorphism is that each new class *completely* overrides methods of its base class. This means that if we apply a method to an object then it is guaranteed that only one method defined in one class in the inheritance hierarchy will be executed (although the decision is made at run-time). The compiler inserts a small piece of intermediate code for each method call which is responsible for method indirection by choosing (dispatching) the method defined for this concrete object type.

In principle, COP could mechanically integrate the object-oriented approach to polymorphism where child concepts completely override parent concepts. However, such a solution would be too artificial and incompatible with the main concept-oriented principles. Therefore COP develops its own mechanism of polymorphism which generalizes the classical one. The main idea is based on the precedence of base reference methods over child reference methods (Principle 2, Section 2.3), which means the ability of parent references to intercept access to child references. Obviously, this means that the parent reference is the very first element in access request processing chain, i.e., it can somehow contribute to the processing of the method call before the target object gets control. Thus an intrinsic feature of this method execution mechanism is that target object methods are not called directly but rather method execution is a sequence of steps.

By intercepting all incoming access requests in the base reference, it is possible to perform some processing and then pass the request to the child reference. (When the request is being processed we can use functions of the parent object.) The child reference gets the access request, processes it and then again passes to the next reference and so on till the target reference. In the simplest case, parent reference does not perform any processing and simply passes it further. And only if it is the last (target) element in the hierarchy, the reference can do some meaningful actions which correspond to the type of this element. For example, let us assume that there is base class `Panel` extended by class `Button`. If we declare a variable of the base class which is then assigned a reference to a button object then method draw applied to this variable will draw a button even though the variable is a panel. In OOP this is done by overriding method draw. In COP, the base reference will intercept the invocation of method draw. Then it can check if there is an extension (a child object) and decide how to proceed. In the case of the button object, the base panel reference can simply pass this request further to the button reference segment which will be responsible for drawing this object of class Button. In other cases, the base panel reference may do something specific to the current level before it passes request to the child element. For example, the panel might fill its background as an intermediate step.

Let us consider another example shown in Listing 7. Concept SavingsAccount is included in concept Account (so one account may have many savings accounts as well as other types of sub-accounts). Both concepts implement method getBalance. The method of Account checks if the child object really exists (line 5) and then either returns its own balance (line 5) or the balance of the child account (line 6). In code we can declare a variable as having base type Account and then the balance returned by getBalance method depends on the real object type. If the object is of concept Account (line 24) then we get one behaviour. If it is of concept SavingsAccount (line 26) then we get different behaviour.

Listing 7. Polymorphism in COP.

```
01 concept Account
02   reference {
03     char[10] accNo;
04     double getBalance() {
05       if(sub == null) return = balance;
06       else return = sub.getBalance();
07     }
08   }
09   object { double balance = 10.0; }
10
11 concept SavingsAccount in Account
12   reference {
13     String subAccNo;
14     double getBalance() {
15       if(sub == null) return = balance;
16       else return = sub.getBalance();
17     }
18   }
19   object { double balance = 20.0; }
20
21 Account account;
22 double balance;
23 account = findAccount(); // Real type is Account
24 balance = account.getBalance(); // = 10.0
25 account = findSavingsAccount();// Type SavingsAccount
26 balance = account.getBalance();// = 20.0
```

Note that `SavingsAccount` assumes that there can be also internal objects (line 15), i.e., it is implemented in the concept-oriented manner where methods are intermediate processing elements getting a request from somewhere and then dispatching them to somewhere for further processing. The polymorphic behaviour is defined by the programmer who writes intermediate methods each contributing to the overall processing. We can include a new child concept in `SavingsAccount` later for example to describe some concrete savings account type and it will be incorporated into the whole access sequence by getting requests from its parent concept.

From this example we see that one and the same method applied to a variable of base type may cause different actions depending on the real type of reference stored in it. In COP, such a method call is a sequence of actions associated with the reference segments. Each intermediate reference and object may contribute to the processing of the access request (as shown in Fig. 4). As a consequence, the real composition of a complex reference influences how requests are processed. In OOP, polymorphism is much simpler and is reduced to choosing the method defined in the real object class which completely overrides its base methods. Thus the method executed by default in OOP is only the last step in a sequence of actions executed in COP. Interestingly, COP does not guarantee that the last method corresponding to the real object type will be reached while in OOP it is always so. The base reference methods override child methods and may finish processing at any moment without continuation. For example, the base method may raise an exception because of security constraints or insufficient resources. Or we could disable method overriding at all. Such an approach is more flexible because request processing is distributed among all constituents at different levels rather than concentrating all the functionality in one class. An advantage of the concept-oriented polymorphism is that the programmer is able to control the whole sequence of access. However, in simple situations it is less efficient because in OOP the indirection used for calling virtual methods is optimized by the compiler.

## 3 Reference Substitution and Resolution

### 3.1 Continuation Method

In the previous sections we have described how the hierarchical structure of references and objects can be modelled using concepts and inclusion relation. Concepts allow for modelling reference structure and behaviour but one of the main questions is how a reference is connected with the represented object, i.e., how meta-transition is implemented and how object methods can be called from its reference. For example, if bank accounts are identified by their numbers then how can we execute methods of the account objects? One solution to this problem is that it is the translator that is responsible for implementing meta-transition.

However, this approach leads to the same problem as that with primitive references in OOP. Namely, the translator is completely unaware of the real scope of the custom references and will use a universal mechanism which is not suitable for many application-specific situations. For example, we might want to implement our accounts as persistent objects and then their state has to be stored in a database or the accounts could be remote objects. The only solution in this case consists in providing a custom meta-transition mechanism for each custom reference which takes into account the purpose of this reference and its specific properties.

In order to provide a mechanism for meta-transition we assume that direct access to objects is provided by primitive references which are substituted by custom references defined by the programmer. Thus *substitution relation* among references is a means of virtualization which binds the virtual world of custom references with real world of primitive references. Substitution and inclusion play crucial role in COP but they have different role and interpretation. Inclusion can be characterized as 'in addition to' this reference where a new more specific reference segment is attached to or concatenated with the parent more general segment. Substitution is interpreted as 'instead of' this reference which means that a virtual reference will be used as a more abstract representation completely substituting this more real reference.

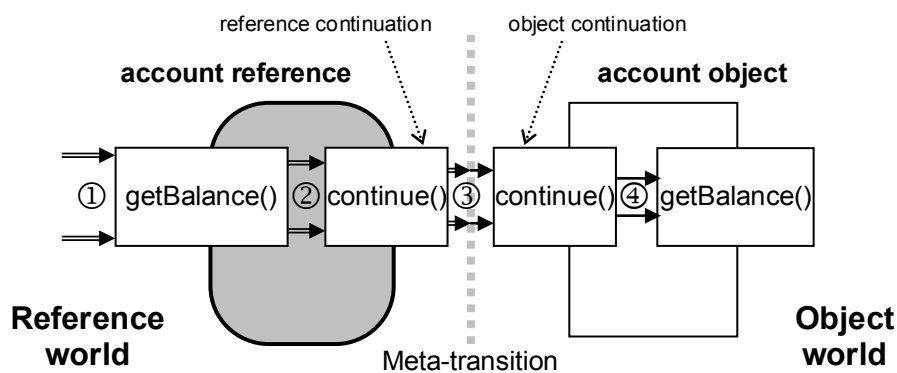

Figure 6. Meta-transition using continuation method.

To resolve custom (virtual) references into the object primitive reference, it is proposed to use a so called *continuation method*. The main task of this method consists in resolving this reference into the substituted primitive reference, i.e., this method is responsible for switching from a reference to the object. Once the primitive reference has been found the task is supposed to be solved, i.e., we assume that the object can be directly accessed and hence the target object method can be called. It is important that continuation method is called automatically whenever it is necessary to cross the border between the identity and entity worlds with the purpose of accessing the object. Thus we still can use objects as if they were directly accessible with the only difference that the continuation method of this concept will be used for access.

The sequence of access using the dual continuation methods is shown in Fig. 6. Notice that in comparison with Fig. 1 it has two continuation methods around the border and between `getBalance` methods. If we apply a method to a reference then the reference method will be executed (step 1). If it calls some object method or otherwise accesses the object then meta-transition is performed implicitly using the continuation method. First, the reference continuation method resolves this reference (step 2) into a primitive reference which again is used for continuation. At this moment the border is intersected (step 3). Then the object continuation method starts automatically from which then the target object method is called (step 4).

Listing 8. Continuation method.

```
01  concept Account
02    reference {
03      char[10] accNo;
04      Object accObject; // Primitive reference
05
06      void continue() {
07        // Resolve account number start access
08        if(accObject == null)
09          accObject = loadAccount(accNo);
10
11        accObject.continue(); // Proceed to the object            (3)
12
13        // Clean up and finish access
14        if( lowMemory() ) {
15          saveAccount(accNo, accObject);
16          accObject = null;
17        }
18      }
19      ...
20    }
21    object {
22      double balance;
23      boolean isAccessed;
24      void continue() {
25        // Enter object and prepare it for access
26        isAccessed = true;
27        continue(); // Proceed to the method                      (4)
28        isAccessed = false;
29        // Clean up and finish access
30      }
31      ...
32    }
```

Listing 8 is an example of the continuation methods which demonstrates the sequence of object access and meta-transition. The continuation method is named `continue` is defined in both the reference class and the object class. The main role of the reference continuation method consists in converting this reference into the substituted primitive reference. In this example we assume that the primitive reference is stored in one of the reference class fields (line 4). If this field is not initialized then we assume that the object state is in some secondary storage and needs to be loaded in memory (line 9). For example, here we might load the state of the account from a database using the account number as a primary key. When the substituted primitive reference is available we can really cross the border and go to the object (line 11). This line is very important because it is rather typical style for concept-oriented programming when no method is called but the necessary actions are executed implicitly. In this case we simply say that we want to continue and the compiler then implements the logic of meta-transition at the level of primitive references. Thus line 11 is where the object continuation method (line 24) starts. This method is guaranteed to be executed before any access and its main role consists in preparing this object for access. In this example we simply set a flag (line 26) which is a signal for other processes that this object is being accessed. When the object is ready for access we again continue (line 27) but in this case the compiler interprets this method call as a point where the target method called explicitly by the programmer can be executed. If the programmer called `getBalance` then it will start at this point. When the target method or any other access request finishes, the process returns to the object continuation method where the access flag is reset (line 28). After that it returns to the reference continuation method where the state of the object can be stored back into the database if necessary (line 15). And finally the process returns to the reference part of the target method such as `getBalance`. Thus the sequence starts from the explicitly called reference method which wraps implicitly called continuation methods (reference and object continuation) which wrap the object method.

## 3.2 Context Stack

Let us assume that there is a complex reference which is used to invoke some method of the represented object. This target object method as well as intermediate methods executed during access can call methods of the parent objects using the 'super' keyword in the same way as it is done in OOP. For example, a method of concept `SavingsAccount` could call methods of its base concept `Account`:

```
concept SavingsAccount in Account
  reference { ... }
  object {
    void getBalance() {
      Person owner = super.getOwner();
      ...
      double limit = super.getCreditLimit();
      ...
      bool isLocked = super.isLocked();
      ...
    }
  }
```

Formally, each access to any object can be performed only via its reference. If we access the parent object then it is represented by the parent segment of the current complex reference and hence this segment has to be resolved for each such method call. In the above example, we would need three resolutions of the parent reference segment for each of these method calls. According to this approach, one access requires one reference resolution and after the access is finished the result of the resolution is lost. Although in many cases like access to the parent object in the above example we know that the result of resolution will be the same, the continuation method will still be executed. If there is many parent method calls and field read/write operations then performance of method execution can be rather low because it requires multiple repeated resolutions of the same reference. Another problem with such a type of access is that frequently we want to guarantee that the same state of the object produced by one resolution procedure is being used.

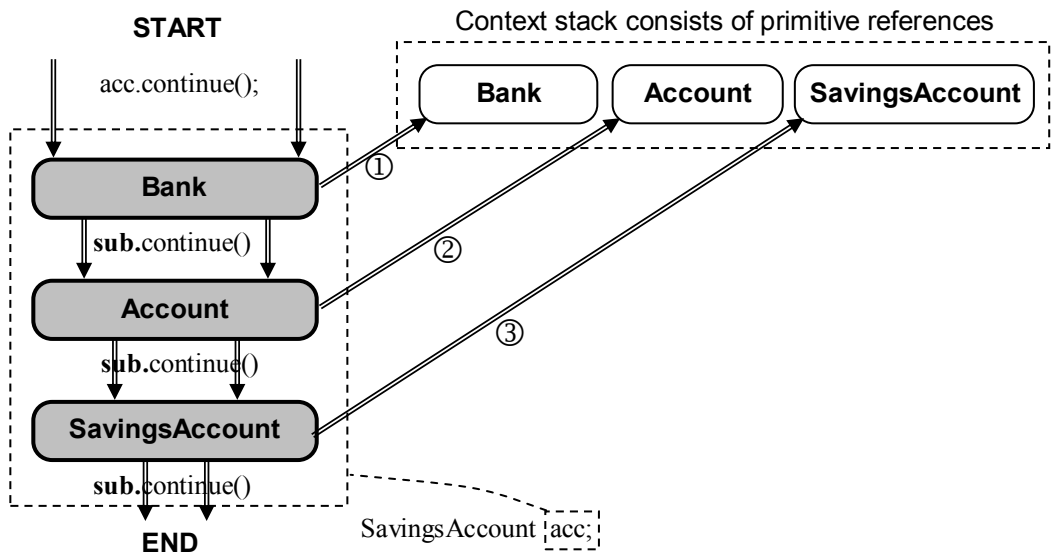

Figure 7. Complex reference resolution and context stack.

This approach is not only inefficient but also not very natural. It is analogous to the situation where we would need to exit and enter again a building in order to talk to several persons within it on different issues. The natural approach would be to enter the building and not to exit before we finish all the necessary actions. In other words, in order to execute several operations in some space, its border has to be crossed (resolved) only once while the intermediate objects should be accessible directly from inside without repeated resolutions. In the case of a complex reference this means that all its segments have to be resolved in advance and the result of the resolution stofred somewhere for future use. After that all the intermediate objects can be accessed directly as many times as needed using the saved primitive references. Thus reference segments are resolved *before* real access takes place and the result of the resolution is stored in a special data structure called *context stack* (Fig. 7). The resolution sequence starts from the first (high) segment and then proceeds to the next segments ending with the last (low) segment. The result of each resolution is pushed on the context stack which grows as each next segment is resolved.

For example, let us assume that an object is represented by a complex reference consisting of three segments of concepts `Bank`, `Account`, and `SavingsAccount`. Initially, just before the access procedure starts, context stack is empty. When the first segment `Bank` is resolved by means of its continuation method (Fig. 7, step 1) it contains a primitive reference to the first object segment of concept `Bank`. Since this moment the bank object is directly accessible. The result of resolving the second segment `Account` is pushed on top of the context stack on the second step, which now contains two primitive references and so on till the last segment (step 3). Finally, the number of elements on it is equal to the number of segments in the complex reference being resolved (3 segments in this example). The top of the context stack is a direct reference to the target object of concept `SavingsAccount`.

Listing 9. Hierarchical access and context stack.

```
01  static Map map = new Map();
02  concept Account
03    reference {
04      char[10] accNo;
05      void continue() {
06        print("=> Account: Resolve");
07        Object o = map.get(this.accNo);
08        o.continue();
09        sub.continue();
10        print("<= Account: Resolve");
11      }
12    }
13    object {
14      double balance;
15      Map map = new Map();
16    }
17
18  concept SavingsAccount in Account
19    reference {
20      String subAccNo;
21      void continue() {
22        print("=> SavingsAccount: Resolve");
23        Object o = super.map.get(this.subAccNo);
24        o.continue();
25        sub.continue();
26        print("<= SavingsAccount: Resolve");
27      }
28    }
29    object {
30      double balance;
31      Map map = new Map();
32    }
```

Let us consider how this sequence of resolution is implemented by continuation methods of concepts in an inclusion hierarchy. The main role of continuation method consists in restoring the primitive reference substituted by this reference. Then it simply passes control to the primitive reference and the process continues on the other side of the border in object continuation method. At this moment the resolved reference is pushed on the stack. However, if there are child objects then they also have to be resolved and hence continuation method has to call them. For example, let us assume that parent concept `Account` has one child concept `SavingsAccount` as shown in Listing 9. Continuation method of concept `Account` resolves its own account number into a primitive reference (line 7) and then passes control further (line 8). Here the mapping from account numbers to objects is stored in a global variable (line 1). Finally reference continuation method calls its child continuation method (line 9) which also has to resolve its reference segment and prepare object for access. If it is `SavingsAccount` then its continuation method resolves sub-account number (line 23) but in this case it uses its parent object for storing the mapping from sub-accounts to objects. Then as usual control is passed to the primitive reference (line 24) and finally this method gives opportunity to a possible child object to make appropriate operations (line 25).

The most important property of this mechanism is that parent objects are directly accessible from their child objects and need not be resolved. So each occurrence of the 'super' keyword in code means *direct* access using the primitive reference from the context stack rather than new resolution via continuation method. Since normally concept functionality is based on using parent concepts, this mechanism leads to significant performance increase because it guarantees that any reference segment is resolved only once for each use of the complex reference.

### 3.3  Creation and Deletion

To demonstrate how objects are represented and accessed we assumed that they already exist. However, before an object can be accessed it needs to be created and this moment is the starting point for its life-cycle. At the end of its life-cycle the object needs to be deleted and after this moment it cannot be accessed anymore. Thus creation and deletion are procedures which limit the life-cycle of any object in time. After creation, a new reference is supposed to be valid which means that it can be used for access. And after deletion this reference is supposed to be invalid which means that it cannot be used for access anymore. One interesting detail in life-cycle management is that creation and deletion deal primarily with references. That is, object existence means presence and validity of its reference (even if the object itself does not yet exist). And vice versa, object non-existence means that its reference is not valid and cannot be used for access (even if the object actually exists).

Object creation and deletion is supposed to be managed by two special methods of concepts named `create` and `delete`. These methods have the same status as another special method – the continuation method – described earlier. In contrast to the continuation methods, creation and deletion methods are normally called explicitly by the programmer. However, they possess all properties of special methods. In particular, these methods are responsible for implementation of meta-transition, i.e., they need to implement some logic of crossing the border from identity to the entity. These methods also have a special meaning for the context stack mechanism.

Listing 10. Creation method for one concept.

```
01 static Map map = new Map();
02 concept Account
03   reference {
04     String accNo;
05     void create() {
06       print("=> Account: Create reference");
07       this.accNo = getUniqueNo();
08       Object o.create(); // Go to object constructor
09       map.add(accNo, o);
10       print("<= Account: Create reference");
11     }
12     void continue() {
13       Object o = map.get(this.accNo);
14       o.continue();
15     }
16     void delete() {
17       print("=> Account: Delete reference");
18       Object o = map.get(this.accNo);
19       o.delete();// Go to object destructor
20       map.remove(accNo);
21       print("<= Account: Delete reference");
22     }
23   }
24   object {
25     double balance;
26     void create() {
27       print("-> Account: Create object");
28       balance = 0;
29       print("<- Account: Create object");
30     }
31     void delete() {
32       print("-> Account: Delete object");
33       balance = 0;
34       print("<- Account: Delete object");
35     }
36   }
37
38   Account account.create();
39
40   $ => Account: Create reference
41   $ -> Account: Create object
42   $ <- Account: Create object
43   $ <= Account: Create reference
```

Just as other methods in COP, creation and deletion methods are dual, i.e., they have two definitions: one in the reference class and one in the object class. Reference creation method is responsible for this reference initialization which means also allocation of the primitive reference it will substitute. Reference deletion method is responsible for deletion of the resources associated with this reference including the substituted primitive references. The main role of object creation and deletion methods coincides with that of constructor and destructor in OOP. In other words, object creation method is intended to initialize the new object just after its creation and before it can be accessed (so it is the very first operation with the object). Object deletion method has to clean it up just before real deletion (so it is the very last operation with the object).

Let us assume that there is one concept `Account` (Listing 10) and we need to define its creation and deletion methods. Creation method generates a unique identifier for the new account (line 7) and then allocates system resources for this object by creating a new primitive reference (line 8). It is precisely the point where the new object is really created. Line 8 is also the point where the object creation method (constructor) is called (line 26). The final step consists in storing the association between this account number (this identity) and the just created primitive reference it is going to substitute. Here we simply store this pair in a global map (line 9). When this object will be accessed this map will be used to resolve object references by the continuation method (lines 13).

Deletion method follows the inverse sequence of steps. It resolves this reference (line 18) and then destroys the restored primitive reference (line 19) by executing object deletion method (destructor) just before real deletion (line 31). After deletion, this reference cannot be used anymore because the account number stored in its field cannot be resolved. Notice also that creation and deletion methods (just as continuation method) do not return any value. Instead, they are applied to a reference with the purpose to initialize or clean up its value. It is also possible define several creation/deletion methods taking some parameters.

In the case of concept hierarchy creation and deletion methods should propagate downwards over the hierarchy in the same way as it is done for continuation method. For example, if concept `Account` has some child concept such as `SavingsAccount` then it could implement its creation method as shown in Listing 11. It is assumed that persons have main accounts with many sub-accounts. Creation method takes one parameter with the name of the owner of the new sub-account. However, it may well happen that this person already has the main account and in this case it has to be reused for creating a new sub-account. Otherwise both a new main account and a new sub-account have to be created. Thus creation procedure checks if an appropriate object can be found (line 6). In the case it is not found a new account number is generated (line 9), new object is created (line 10) and the association between them is stored (line 11). (Alternatively, we might simply call creation method without parameters as implemented in the previous example.) If the main account is found for this user then we resolve this reference by calling continuation method (line 13). Finally, we call creation method of the child reference (line 14) so that if it is a savings account then it will be created. Notice that when a child creation method is executed, its parent already exists (either new or reused) and is directly accessible because its primitive reference is in context stack.

Listing 11. Creation or reuse of available objects.

```
01 static Map map = new Map();
02 concept Account
03   reference {
04     String accNo;
05     void create(String name) {
06       this.accNo = findAccount(name);
07       Object o;
08       if(this.accNo == null) {
09         this.accNo = getUniqueNo();
10         o.create();
11         map.add(accNo, o);
12       }
13       else o.continue();
14       sub.create();
15
16     }
17     void continue() {
18       Object o = map.get(this.accNo);
19       o.continue();
20       sub.continue();
21     }
22   }
23   object {
24     double balance;
25     Map map;
26     void create() {
27       balance = 0; map = new Map();
28     }
29   }
```

Thus it is not necessary to really create all constituents of a complex object and each creation method can choose its own logic of creation which is appropriate for this problem domain. In particular, it is possible to implement lazy creation when we generate only a unique reference while real object creation will be performed only when this object is accessed. Another use case is where concept maintains a pool of objects as a list of primitive references. When a new object is requested to be created a primitive reference is taken from this pool rather than allocated by the system routine. Deletion method could simply mark the deleted primitive reference as unused and return it to the pool. It is also possible to assign explicitly some parent segment with valid values pointing to an existing object (such as main account). Then creation procedure will interpret it as a request to initialize only last segments. There exist also many other applications of this life-cycle management mechanism and it is especially useful for tasks where some complex logic is needed.

## 4 Operations with References

References are passed and stored by-value in variables, fields, method parameters, return value and other elements of the program which are supposed to have an explicitly declared type as some concept. This declared type is assigned to a variable and restricts possible references that it can store. The real type of a reference stored in the variable can vary but it always has at least this declared type. For example, if we declare a new variable as having type `Account`

```
Account account;
```

then it can store a reference of type `Account` or `SavingsAccount` but not `Bank`.

The declared type of a variable is returned by the operator `concept()` while the real type of the reference stored in it is returned by the operator `instanceof()`. If we assume that '<' (less than) denotes 'child-parent' relation between concepts, i.e., child concepts are less than the parent concept, then for any variable the following inequality holds:

```
concept(var) >= instanceof(var)
```

This means that any variable stores a reference of its declared concept or some of its child concept.

The fact that real references can be longer than the declared type of the variable is analogous to the fact that variables can store more specific objects in OOP. Thus 'longer reference' is equivalent to 'more specific reference'. A new feature of COP is that real references can be made shorter by cutting off some starting (higher) segments. For that purpose we use the notion of context which is defined as follows:

**Definition 4 [Context of a reference].** *Context* is the parent concept of the first segment of a reference.

If a reference is divided into implicit part (higher segments) and explicit part (lower segments stored in the variable) then context is the real type of the implicit part. For example, if the first segment is of type `Account` then its context is `Bank` and if the first segment is of type `SavingsAccount` then its context is `Account`. The real context of a reference is returned by the operator `contextof()`, i.e., this operator returns the parent concept of the first segment of its parameter. In order for a reference not to be empty the following condition has to be always satisfied:

```
contextof(var) > instanceof(var)
```

By default all references have full context, i.e., all parent segments are included in the reference and in this case `context(var)==Root`. If it is necessary to restrict context for some variable then it can be done by specifying the necessary context along with the type of the variable as follows:

```
Account : SavingsAccount savAcc;
```

Here we use two types separated by colon where the first type declares the context while the second type is normal type. This declaration means that a reference stored in this variable must be at least `SavingsAccount` and its representation will start from `Account`. To get the declared context of a variable the operator `context()` will be used.

Above we demonstrated how colon operator can be used to impose constraint on reference type when declaring a variable. The same operator can be applied to variables themselves and in this case it is interpreted as concatenation of references. The first parameter (before colon) provides higher segments while the second parameter provides lower segments. For example, if one variable stores a bank reference while another variable stores a reference to an account (without bank) then we can create a fully qualified account reference as follows:

```
Bank bank = getBank(); // Only bank
Bank : Account account = getAccount(); // Only account
Account account = bank : account; // Account in bank
double balance = (bank : account).getBalance();
```

The third line here stores a fully qualified reference to the bank account while the fourth line concatenates two references and applies a method to the obtained fully qualified reference.

Given a reference we can change its context (first segment) and type (last segment) by using left and right cast operations. Left cast is used to specify the necessary context and results in either adding new segments (if new context is larger than the existing) or removing some starting segments (if new context is less than the existing). Left casting is written as the concatenation of the context specified as some concept name and the variable storing the reference:

```
LeftConcept : var
```

Here the type of context is written on the left and is separated from the variable by colon. The reference returned by this operator has the context specified in the argument:

```
contextof(LeftConcept : var) == LeftConcept
```

One important use of left casting consists in converting a reference to a global reference by adding to it the maximal context. For example, if we have only savings account number but need to pass it as a fully qualified account description then we attach the global context as follows:

```
fullAccountRef = Root : savingsAccount;
```

Left casting can be also used to shorten a reference. For example, if we have an account variable and want to get only a sub-account reference stored in it (without the main account part) then it can be done as follows:

```
subAccountRef = conceptof(account) : account;
```

The operation of right casting changes the real type of this reference by either removing some last segments or by adding new (empty) last segments. We write the desired right segment concept after the reference separated by colon:

```
var : RightConcept
```

This operation results in a new reference with the real type equal to the specified concept name:

```
instanceof(var : RightConcept) == RightConcept
```

Right casting can be used to extract starting segments cutting off its tail. For example, if we have an account reference and need to extract only the main account identifier then it can be done as follows:

```
mainAcountRef = account : conceptof(account);
```

In this way it is also possible to get a reference to any intermediate object represented by this reference by specifying its concept.

## 5 Related Work

The role of identities has probably never been underestimated in computer science but most existing approaches to identity modelling have a theoretical nature and propose conceptual or logical foundations for dealing with identities and identification. We do not know any work that would proceed from the assumption that identities have their own behaviour and are responsible for a great deal of the system functionality as it is assumed in COP.

Very interesting model where objects and their identities are connected within one model was proposed by Kent [12]. In this model, scope is analogous to what we mean by space or context, i.e., it is assumed that everything exists in some scope. However, this scope is not a normal element – it is an additional construct in the model. In contrast, in the concept-oriented model context is just a normal element, i.e., any element exists in some context and any context is an element. Scope in Kent's model has the same role with respect to references as in our model: "A token belongs to a scope, which determines its status and meaning as a reference, based on the status and meaning of its corresponding symbol in that scope." However, this role is assigned in an informal manner while in our approach it is a concrete mechanism implemented using special functions and obeying a concrete sequence of access. Another feature of the Kent's model is that scopes are not intrinsically nested. Strictly speaking, it is noticed that scopes can be nested: "Scopes could be nested, but that's beyond the concern of this paper." Yet the nesting is not used as a high level principle while in the concept-oriented model this relation is of primary importance, i.e., we assume that elements cannot exist without a hierarchy and without a parent element (context). The next property of the Kent's model is that it does not have substitution relation among elements, i.e., objects have identities but it is not described how these identities are used for access. This means that identities have a mechanism of access as their intrinsic or primitive property. In our approach substitution relation is one the corner stones. It should be noticed however that Kent's model allows for synonyms.

Another interesting model is proposed in [36]. The authors provide a definition of an object identification schema, study its properties and compare with the existing identification mechanisms such as oids, keys and surrogates. In great extent, this paper consolidates what was known at that time about object identification. However, it does not integrate the theoretical conceptual analysis of the role, applicability and limits of object identification into any practical framework or setting and therefore can be viewed as a conceptual vision.

The only known approach for directly modelling references consists in using *smart pointers* in C++ [34]. The idea is that conventional class can be used as a reference class while its parameter is the represented object class. For example, a new variable could be defined as follows:

```
AccountRef<Account> account;
```

Here `AccountRef` is a normal class which however is intended for representing other objects with the class specified as the template parameter. If we provide `Account` as a parameter then this variable will represent objects of class `Account` using a reference of class `AccountRef`. The same idea of having a reference class parameterized by a target object class is implemented in the Transframe programming language [31] where a variable with a custom reference is defined as follows:

```
account: AccountRef of Account;
```

However, this technique is actually a programming pattern where conventional classes are adapted for the purposes of object identification. In particular, this approach has the following problems and limitations:

- Smart pointers assume that any new variable requires two parameters: a reference class name and an object class name. These two classes are defined separately and are completely unaware of each other because they are paired only when a new variable is declared. In COP the association between reference class and object class is established within one concept and variables are declared precisely as in OOP using only one concept name.

- Smart pointers can be viewed as proxies because they are created as instances of conventional classes which are then used instead of the target objects. In COP we have the illusion of working directly with the target object while the indirection is completely hidden.

- Smart pointers allow the programmer to implement simple substitutes passed by value but it is difficult to implement hierarchical (complex) references consisting of several segments. And it is even more difficult to implement objects consisting of several separate segments.

- Smart pointers intercept method calls by overloading access operator. In COP the interception of individual methods is performed by reference methods of the concept which have precedence over object methods and have the opposite overriding direction. General interception can be also implemented via continuation method.

As already mentioned, references with their structure and behaviour can be modelled only by smart pointers in C++ and referential classes in Transframe. However, there exist numerous approaches for modelling indirection and intermediate behaviour. Probably the simplest approach to injecting intermediate functions consists in applying the Proxy pattern [5]. Proxy is a conventional class which is intended to emulate the interface of the corresponding target class but inserts some intermediate functionality. For example, if it is necessary to do something before an account object is accessed we can define a proxy class, called `AccountProxy`, and then use it instead of the class `Account`. Since we explicitly use the proxy in the program, all references will point to this proxy and hence it is not true interception or injection but rather can be qualified as manual indirection. This technique has the following obvious problems and limitations:

- If the target class changes then its proxies need to be updated manually because the relation between proxy and the target object is maintained only by the programmer.

- Proxy is developed for one target class because its task consists in explicitly simulating its behaviour. It is difficult to develop a kind of generic proxy which could be used for many different target classes such as a security proxy for authorizing access. In COP it is done by describing a general-purpose reference in a parent concept and then including other concepts in it so that one reference will be used for representing different types of objects.

- It is difficult to impose behaviour in a nested manner (creating a proxy for a proxy) because it requires even more manual support and such a program is even more sensitive to changes which have to be propagated all over the source code.

- This approach allows the programmer to implement indirect access but it does not provide means for modelling references which are passed and stored by value instead of native references. A reference to a proxy is still a normal native reference.

- Proxies do not guarantee that the represented object cannot be accessed directly (without proxy) because it has an independent class which can be instantiated. COP gives more protection because concept instantiation procedure returns a reference in the format defined in this concept.

One of the most interesting approaches to programming is aspect-oriented programming (AOP) [15]. In terms of COP, an aspect is a programming construct that modularizes intermediate functionality (in advices). Aspects describe intermediate functionality (and data) injected into special points in the program (join-points) which are specified by means of regular expressions. What is similar between AOP and COP is that method invocations are inherently indirect and can trigger quite complex intermediate actions. Another similarity is that these intermediate actions can be effectively modularized and therefore AOP and COP can be viewed as two alternative approaches for separation of (cross-cutting) concerns as formulated in [4]. AOP uses aspects in addition to classes for this purpose while COP modularizes cross-cutting functionality in parent concepts which then inject these functions into child concepts by intercepting incoming requests. Thus two features of COP make aspect-orientation possible: reverse method overriding and set-orientation. Reverse method overriding allows parent concepts to intercept accesses to their child concepts while set-orientation means that one parent concept modularizes (cross-cutting) behaviour which is common for many child concepts. The most important difference between AOP and COP is the direction of dependence between the injecting module and the target points which are being modified. More specifically, aspects know explicitly the points where the intermediate functions will be injected and the target join-points do not know what other code will modify their behaviour (Fig. 8, left). In COP this dependence has the opposite direction. Namely, the injecting module (parent concept) is unaware of the points where its code will be used (Fig. 8, right).

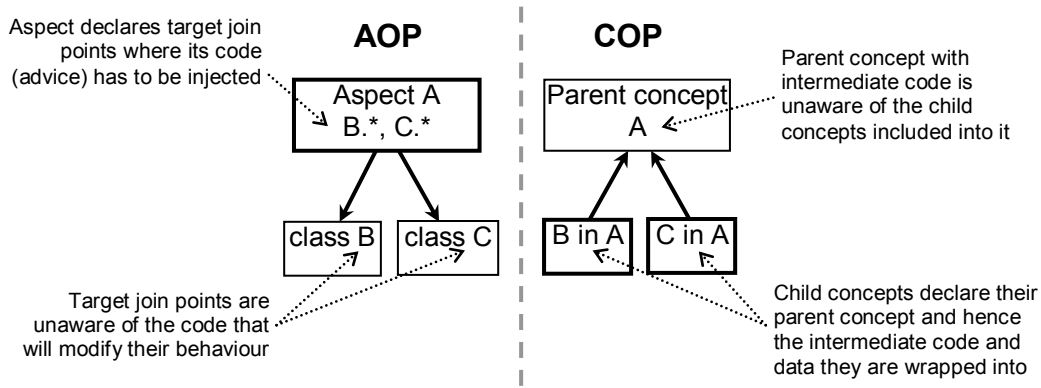

Figure 8. Aspect-oriented programming vs. concept-oriented programming.

The mechanism of dual methods in COP is similar to *super/inner* methods of classes [7] which is implemented in the Beta programming language [16, 17, 18]. In particular, the inner methods are designed in such a way that they implement the same sequence of access as that in reference methods. However, the mechanism of super/inner methods is implemented as an addition to normal classes. Hence it can be viewed as an enhancement to OOP aimed at providing means for object protection from outside. In COP, this behaviour is implemented using a completely different approach, namely, by means of concepts.

The structure of elements produced by inclusion relation in COP is the same as in prototype-based languages [Lie86]. The difference between them is that prototype-based programming implements the mechanism of sharing behaviour via delegation [33] without using classes while COP relies on both mechanisms of inheritance (class hierarchy and instance hierarchy). Essentially, COP combines class-based and prototype-based approaches and makes the structure symmetric because both classes (concepts) and instances exist in a hierarchy. In COP, a parent element is a prototype but on the other hand, any element is an instance of some concept. The hierarchy in COP is built from pieces which is similar to the split object model [1]. The difference is that it is based on identities where each child has a relative reference with respect to its parent. Another difference is that elements in COP use two-way delegation by forwarding request either to the parent (using super keyword) or to the child (using sub keyword) while prototype-based languages use traditional delegation where a message is sent to the prototype for the default processing.

The concept-oriented approach relates also to so called *context-oriented* methods which are aimed at bringing context dependence into programming [3, 6, 11, 22]. These methods introduce languages constructs and mechanisms which allow the programmer to put objects in a context changing their behaviour at run-time. For example, in the ContextL programming language it is done by means of the keyword 'in-layer' while in COP we use 'in' which generalizes inheritance and 'super' to access the context.

There exist also many other approaches which can be used for modelling indirection and intermediate functions like meta-object protocol [13, 14], mixins [2, 32], language-oriented programming [35]. However, their fundamental assumptions and proposed methods are completely different from those made in COP. It should be noted that there exist also approaches to programming having the same name but which are actually based on very different notions and do not relate to our work. For example, B. McConnell proposed an approach to programming which is also called concept-oriented programming [20]. This method uses DNS-like system which extends object-oriented programming languages. This system stores reusable code that can be loaded from programs as a library. An advantage is that programmers can write tiny programs that load these libraries. The mechanism of concepts also exists in generic programming where it is used to describe a set of operations supported by a type [8].

There can be two approaches to programming based on using concepts defined as a pair of one reference class and one object class. These approaches, called COP-I and COP-II, depend on the role played by these classes and their responsibilities. COP-I [23] assumes that references of a concept represent objects of its *child* concepts (not this concept). For example, if we define concept `Account` in COP-I then its references are intended to represent child objects such as `SavingsAccount` or `CheckingAccount`. How then objects of this concept are represented? They are represented by references of the parent concept. For example, account objects could be represented by references of concept `Bank`. Thus concepts in COP-I describe one space object with a set of internal references for representing internal objects. This space object knows references of its internal objects but does not know what kind of objects they will represent. What is important is that an object is responsible for managing a *set* of references.

COP-II [27, 28, 30] described in this paper assumes that references of a concept represent objects of this same concept. Thus concepts in COP-II describe one element which consists of one object and one reference. Both approaches have some advantages and disadvantages. COP-I has some very attractive features but later on when trying to remove some subtle problems we gradually switched to the second approach described in this paper. Yet, it is still a challenging problem to decide which of these two approaches is more natural.

One of the most interesting features of COP is that it is essentially part of a data model, which is called the concept-oriented model of data (COM) [24, 25, 26, 29]. Moreover, COP has been developed to be an integral part of this data model and to serve as the basis for a new data modelling approach. COM, which is based on the theory of ordered sets, can be viewed as an extension of COP where data semantics is added into the programming model. Such integration of programming and data modelling allows us to decrease the problem which is known as impedance mismatch. In particular, data elements in COM are also supposed to consist of two parts – identity and entity. But in addition to identity modelling based on inclusion relation we can model data semantics using the theory of ordered sets and such operations as projection and de-projection.

# 6 Conclusions

In this paper we have described a new approach to programming the main new feature of which is that it brings references into the focus of computer programming. References in COP are made as important as objects in OOP and together, in the form of reference-object couples, they constitute the main building block of any program. References in COP are not simply special objects but rather they have a very specific role which is dual to that of objects. The shift of paradigm here is that any system has two types of functionality – explicit object business methods and implicit intermediate methods – which are separated in a principled manner in COP as two concerns. Any object is represented and accessed indirectly using its reference and some activity is being always executed implicitly during object access. In particular, this ability of references to implicitly intervene into the process of program execution allows us to solve the problem of separation of concerns as it is formulated in AOP but using different basic postulates. COP assumes that the cross-cutting behaviour is precisely what references are responsible for and hence we can modularize these functions in reference classes rather than in aspects. In this sense COP is an interesting alternative to AOP.

COP would not be so interesting approach if we simply introduced references and reference classes into programming. Its real advantages come from the notion of concept as a new programming construct which unites object classes and reference classes into one whole. The shift of paradigm here is that now references and objects are considered two parts of one and the same thing and hence we have to model this thing as a primary element by distributing its functionality between object part and reference part according to the application requirements. This separation is made only within concepts when they are defined while the program still manipulates concept instances abstracting from their division into objects and references. An informal analogy here is the use of complex numbers in mathematics which have two constituents but are still manipulated as a whole.

An important advantage of COP is that it generalizes OOP, i.e., we can add new features gradually where necessary. In particular, concept is a generalization of class and we can convert classes into concepts by adding the corresponding features to them. And concept inclusion relation is a generalization of inheritance which allows us to have classical inheritance in one design and (automatically) switch to inclusion if we provide some additional concept-oriented features. Taking into account that COP generalizes OOP, allows for separation of concerns in AOP style, is part of the concept-oriented data model by decreasing impedance mismatch and provides many other interesting features it can be rather perspective direction for further research and development activities in the area of programming languages and system design.